\documentclass[fleqn,10pt]{wlscirep}
\usepackage[utf8]{inputenc}
\usepackage[T1]{fontenc}
\usepackage{amsfonts, amsmath, amssymb, amsthm, color, bm}
\usepackage{bbold}
\usepackage{braket}
\usepackage{multicol, multirow}

\newcommand{\etal}{\textit{et al.~}}

\newcommand{\rev}[1]{#1}

\title{Constrained Quantum Optimization for Extractive Summarization on a Trapped-ion Quantum Computer}
\author[1,2,3,+]{Pradeep Niroula}
\author[1,+, *]{Ruslan Shaydulin}
\author[1,+]{Romina Yalovetzky}
\author[1]{Pierre Minssen}
\author[1]{Dylan Herman}
\author[1]{Shaohan Hu}
\author[1]{Marco Pistoia}

\affil[1]{JPMorgan Chase, New York, NY, USA}
\affil[2]{Joint Center for Quantum Information and Computer Science, NIST/University of Maryland, College Park, MD, USA}
\affil[3]{Joint Quantum Institute, University of Maryland, College Park, MD, USA}
\affil[+]{These authors contributed equally.}
\affil[*]{ruslan.shaydulin@jpmchase.com}

\date{\today}

\let\vec\bm

\newcommand{\Gate}[1]{\textsc{#1}}

\newcommand{\zgate}{\Gate{z}}
\newcommand{\ygate}{\Gate{y}}
\newcommand{\xgate}{\Gate{x}}

\newcommand{\cnotgate}{\Gate{cnot}}

\newcommand{\F}{\mathcal{F}}

\begin{abstract}
Realizing the potential of near-term quantum computers to solve industry-relevant constrained-optimization problems is a promising path to quantum advantage. In this work, we consider the extractive summarization constrained-optimization problem and demonstrate the largest-to-date execution of a quantum optimization algorithm that natively preserves constraints on quantum hardware. We report results with the Quantum Alternating Operator Ansatz algorithm with a Hamming-weight-preserving XY mixer (XY-QAOA) on trapped-ion quantum computer. We successfully execute XY-QAOA circuits that restrict the quantum evolution to the in-constraint subspace, using up to 20 qubits and a two-qubit gate depth of up to 159. We demonstrate the necessity of directly encoding the constraints into the quantum circuit by showing the trade-off between the in-constraint probability and the quality of the solution that is implicit if unconstrained quantum optimization methods are used. We show that this trade-off makes choosing good parameters difficult in general.  We compare XY-QAOA to the Layer Variational Quantum Eigensolver algorithm, which has a highly expressive constant-depth circuit, and the Quantum Approximate Optimization Algorithm. We discuss the respective trade-offs of the algorithms and implications for their execution on near-term quantum hardware.
\end{abstract}
\begin{document}

\flushbottom
\maketitle

\thispagestyle{empty}

\section{Introduction \label{sec:intro}}

Recent advances in quantum hardware~\cite{Arute2019,Wu2021,Madsen2022} open the path for practical applications of quantum algorithms. A particularly promising target application domain is combinatorial optimization. Problems in this space are prominent in many industrial sectors, such as logistics \cite{sbihi2007combinatorial}, supply-chain design \cite{eskandarpour2015sustainable}, drug discovery \cite{kennedy2008application}, and finance \cite{soler2017survey}. However, many of the most promising quantum algorithms for optimization are heuristic and lack provable performance guarantees. Moreover, limited capabilities of near-term quantum computers further constrain the power of such algorithms to address practically-relevant problems. Therefore, it is crucial to perform thorough evaluations of promising quantum algorithms on state-of-the-art hardware to assess their potential to provide quantum advantage in optimization.

When conducting such evaluations, the choice of the target optimization problem is of particular importance. Many well-studied theoretical problems---such as maximum cut (MaxCut)~\cite{wang2018quantum,zhou2020quantum,crooks2018performance,shaydulin2021qaoakit,Wurtz2021Bounds,harrigan2021quantum,shaydulin2021error}, maximum independent set~\cite{ebadi2022quantum} and maximum $k$-colorable subgraph~\cite{bravyi2022hybrid}---have been used to evaluate the performance of quantum optimization algorithms. These problems have several advantages: they are well-characterized theoretically, have strong hardness guarantees, and are easy to map to quantum hardware. At the same time, however, they do not correspond directly to the practically-relevant problems that are solved daily in an industrial setting. 

In this work, we consider the problem of \textit{text summarization}, 
where the goal is to generate a shortened representation of an input document without altering its original meaning. 
This process is commonly used to produce summaries of news articles \cite{filippova2009company} and voluminous legal documents \cite{bhattacharya2021incorporating}. 
Specifically, we focus on a version of text summarization known as \textit{extractive summarization} (ES), 
wherein the summary is produced by selecting sentences verbatim from the original text. 
In our experiments,
we map the ES problem to an $\mathsf{NP}$-hard constrained-optimization problem using the formulation introduced by McDonald~\cite{McDonald}. The resulting optimization problem is then solved on a quantum computer. We impose a constraint on the number of sentences in the summary, which is enforced using a penalty term in the objective, or natively by limiting the quantum evolution to a constraint-restricted subspace.

ES is a particularly interesting problem to consider since it has challenges that are similar to those of many other industrially-relevant use cases. First, it is constrained, making it necessary to either restrict the quantum evolution to the corresponding subspace, or introduce large penalty terms into the formulation. Second, it lacks simple structures, such as symmetries~\cite{Shaydulin2012}. Third, unlike commonly considered toy problems, such as MaxCut, the coefficients in its objective are not necessarily integers, which can make the optimization of quantum algorithm parameters hard~\cite{2201.11785}.

In this paper, we present experimental and numerical results demonstrating the challenges associated with solving constrained-optimization problems with near-term quantum computers. Our contribution is twofold. 

First, we demonstrate experimental results showing successful execution of the Quantum Alternating Operator Ansatz algorithm with a Hamming-weight-preserving XY mixer (XY-QAOA)~\cite{2019aoa,Wang2020} on the quantum processor Quantinuum H1-1. We use all 20 qubits of H1-1 and execute circuits with two-qubit gate depth of up to 159 and two-qubit gate count of up to 765, which is the largest such demonstration to date. We additionally report results from the execution of the Layer Variational Quantum Eigensolver (L-VQE)~\cite{Liu2022}, which is a recently-introduced hardware-efficient variational algorithm for optimization. We obtain approximation ratios of up to $92.1\%$ and in-constraint probability of up \rev{$91.4\%$} on the H1-1 device.

Second, we motivate our algorithm choice by highlighting the trade-off between the quality of the solution and the in-constraint probability which is implicit in applying unconstrained near-term quantum optimization algorithms to constrained problems. This trade-off suggests the need to carefully engineer the parameter optimization strategy, which is difficult in general. We show how this trade-off can be avoided by either using a sufficiently expressive circuits such as L-VQE (at the cost of the increased difficulty of parameter optimization) or, more naturally, by encoding the constraints directly into the circuits as in the case of XY-QAOA.

The remainder of this paper is organized as follows. Section~\ref{sec:problem} introduces the quantum algorithms used. Section \ref{sec:ESasQUBO} describes how the ES problem is formulated as an optimization problem. Section \ref{sec:methods} details the methodology we adopted to generate the optimization problem instances and solve them on real quantum hardware. Section \ref{sec:expResults} illustrates the experimental results we obtained by executing the algorithms and discusses the advantages and downsides of them. Section~\ref{sec:related} discusses previous hardware demonstrations. Finally, Section \ref{sec:Discussion} summarizes our findings and their significance.
Additional technical details on the algorithms, hyperparameter and ES problem are presented in the Appendix at the end of the paper.

\section{Problem Description \label{sec:problem}}

For a given objective function $f$ defined on the $N$-dimensional Boolean cube \rev{and a set of feasible solutions $\F\subseteq\{0,1\}^N$}, consider the problem of finding a binary string $\vec{x}\in\rev{\F}$ that maximizes it:
\begin{align}
    \max_{\vec{x} \in\rev{\F}} f(\vec{x}).
    \label{eq:opt_unconstrained}
\end{align}
\rev{The set of feasible solutions $\F$ is typically given by constraints of the form $g(\vec{x})=0$ or $g(\vec{x})\leq 0$. A binary string $\vec{x}\in\F$ is said to be ``in-constraint''.} Let $C\in\mathbb{C}^{{2^N} \times {2^N}}$ denote the Hamiltonian (Hermitian operator) encoding $f$ on qubits. This operator is diagonal in the computational basis ($C = \text{diag}(f(\vec{x}))$) and is defined by its action on the computational basis: $C\ket{\vec{x}} = f(\vec{x})\ket{\vec{x}}, \forall\vec{x}\in \{0,1\}^N$. 

QAOA~\cite{farhi2014quantum,Hogg2000} solves the problem \eqref{eq:opt_unconstrained} by preparing a parameterized quantum state
\begin{equation}
    \prod_{j=1}^p \left[e^{-i\beta_j \sum_{k=1}^N \xgate_k}e^{-i\gamma_j C} \right]\ket{+}^{\otimes N},
\end{equation}
where $\xgate_j$ denotes a single-qubit Pauli $\xgate$ acting on qubit $j$ and the initial state $\ket{+}^{\otimes N}$ is a uniform superposition over all computational basis states. The parameters $\vec{\beta}, \vec{\gamma}$ are chosen using a classical algorithm, typically an optimization routine~\cite{shaydulin2019multistart}, with the goal of maximizing the expected objective value of QAOA state-measurement outcomes. The depth of a QAOA circuit is controlled by a free parameter $p$. In the limit $p\rightarrow\infty$, QAOA can solve the problem exactly via adiabatic evolution~\cite{farhi2014quantum}. Additionally, there exist lower bounds on the performance of QAOA when solving MaxCut in finite depth~\cite{Wurtz2021Bounds,wurtz2021fixedangle}, and QAOA achieves performance competitive with best-known classical algorithms on the Sherrington-Kirkpatrick  model \cite{Sherrington-Kirkpatrick} in infinite-size limit and finite depth~\cite{Farhi2020SK,2110.14206}.

In the remainder of this paper, XY-QAOA refers to the Quantum Alternating Operator Ansatz algorithm with a Hamming-weight-preserving XY mixer~\cite{2019aoa}, whereas QAOA indicates the Quantum Approximate Optimization Algorithm~\cite{farhi2014quantum}.

L-VQE~\cite{Liu2022} solves problem \eqref{eq:opt_unconstrained} by preparing a parameterized state:
\begin{equation}
    \prod_{j=1}^p \big[U(\vec{\theta}_j)\big]V(\vec{\theta}_0)\ket{0},
\end{equation}
where $U$ is a circuit composed of linear nearest-neighbor \cnotgate~gates and single-qubit rotations, $V$ is a tensor product of single-qubit rotations and $\ket{0}$ is the $N$-qubit vacuum state. Due to the structure of the circuit, the two-qubit gate depth is very low ($4 \times p$ for a circuit with $p$ layers). We refer interested readers to Ref.~\cite{Liu2022} for the precise definition of the circuit.

While QAOA and L-VQE can tackle constrained-optimization problems in which the constraint is enforced by a penalty term, the output of these algorithms is not guaranteed to satisfy the constraint. Moreover, a penalty term introduces a trade-off between the in-constraint probability and the quality of the in-constraint solution, as discussed in Section~\ref{sec:qaoa_tradeoff}. The Quantum Alternating Operator Ansatz algorithm~\cite{2019aoa} overcomes this limitation by using a parameterized circuit which limits the quantum evolution to a constraint-preserving subspace. The general form of this circuit is given by
\begin{equation}
        \prod_{j=1}^p \left[U_M(\beta_j)U_C(\gamma_j) \right]\ket{s},
\end{equation}
where $U_C$ is the \textit{phase operator}, which encodes the objective and is diagonal in the computational basis, $U_M$ is a non-diagonal \textit{mixer operator}, and $\ket{s}$ is some initial state. QAOA circuit can be recovered as a special case by setting $U_M(\beta_j) = e^{-i\beta_j \sum_{k=1}^N \xgate_k}$, $U_C(\gamma_j)=e^{-i\gamma_j C}$ and $\ket{s}=\ket{+}^{\otimes N}$. In this paper, we focus in particular on the Hamming-weight constraint of the form $\sum_{j=1}^N x_j = M$, which in our case corresponds to fixing the size of the summary. While many variations of this algorithm exist~\cite{2019aoa}, we only consider the XY-QAOA version, which lends itself well to implementation on near-term hardware. We let the initial state be the uniform superposition over all binary strings with Hamming weight $M$,  $U_C(\gamma_j)=e^{-i\gamma_j C}$ and $U_{M}^{\xgate\ygate}(\beta_j) = \prod_{k=1}^N e^{-i\frac{\beta_j}{2}(\xgate_k\xgate_{k+1}+\ygate_k\ygate_{k+1})}$~\cite{Wang2020}.
Given that $\sum_{k=1}^N\xgate_k\xgate_{k+1}+\ygate_k\ygate_{k+1}$ commutes with $\sum_{k=1}^N\zgate_i$, the evolution produced by the resulting circuit is restricted to the span of computational basis states with Hamming weight $M$, as desired.

\subsection{Extractive Summarization as an Optimization Problem \label{sec:ESasQUBO}}

Extractive summarization (ES) is an interesting problem to evaluate the performance of quantum optimization algorithms due to its practical importance and complex structure. The goal of ES is to pick a subset of the sentences present in a large document to form a smaller document such that this subset preserves the information content in the original document, i.e., summarizes it. While many approaches to solving ES exist (see e.g. \cite{xu2019discourse, zhong2020extractive, liu2019fine, list-summarization-methods}), in this work we focus on a particular formulation of ES as an optimization problem. Specifically, we consider the problem of \textit{maximizing the centrality} and \textit{minimizing the redundancy} of the sentences in the summary under the constraint that the  total  number of sentences in the summary is fixed. This formulation of ES has been proposed by R. McDonald and shown to be $\mathsf{NP}$-hard to solve exactly~\cite{McDonald}.

We now introduce the necessary notation and formally define the problem. An ES algorithm maps a document of $N$ sentences to a summary of $M < N$ sentences. Let the sentences be denoted by integers $i \in [N] := \{0, 1, \dots N-1\}$ according to the order in which they appear in the document. An extractive summary is a vector $\vec{x} \in \{0, 1\}^N$, where $x_i = 1$ if and only if sentence $i$ is included in the summary text associated with $\vec{x}$. The summary should identify sentences that are central, meaning important, to the document. The salience of a sentence is measured with some centrality which is a map $\mu : [N] \xrightarrow[]{} \mathbb{R}$ satisfying the following property: $\mu(i) > \mu(j)$ if and only if sentence $i$ contains more information about the document than $j$. At the same time, to keep the summary short, it is desirable to ensure that the sentences in the summary are not redundant. The overlap in information content between two sentences is measured with \textit{pairwise similarity} which is a symmetric map $\beta: [N] \times [N] \xrightarrow[]{} \mathbb{R}$ that satisfies the following property: $\beta(i, j) > \beta(i, k)$ if and only if sentence $j$ is more similar to $i$ than $k$ is to sentence $i$. We discuss the particular choices of measures of sentence centrality and pairwise similarity in Appendix \ref{sec:appendix-formulation}. 

The extractive summarization is formulated as the following optimization problem~\cite{McDonald}:
\begin{equation}
\small
\begin{split}
    \max_{\vec{x} \in \{0,1\}^N } &\sum_{i=0}^{N-1} \mu(i)x_i - \lambda \sum_{i\neq j} \beta(i, j) x_i x_j,  \\
    \text{s.t.} & \sum_{i=0}^{N-1} x_i = M
\end{split}
\label{eq:summary-optimization}
\end{equation}
where the parameter $\lambda$ controls how the inclusion of similar sentences in the summary is penalized. The objective of this maximization problem is to increase the information content of the sentences in the summary while ensuring that the total pairwise similarity between sentences is low, relative to $\lambda$. Refer to Appendix \ref{appendix:hyperparameter} that shows how this parameter affects the quality of the summaries obtained.

This problem can be solved directly by a quantum algorithm that can preserve constraints. On the other hand, to solve this problem using an unconstrained-optimization algorithm, we must convert this problem to an unconstrained one by adding a penalty term to enforce the constraint. The penalty term is minimized when exactly $M$ sentences are selected. This term is weighted with the parameter $\Gamma$. Including the constraint on the lengths of summaries, the optimization problem becomes the following:
:
\begin{equation}
\small
\begin{split}
    \max_{\vec{x} \in \{0,1\}^N } \sum_{i=0}^{N-1} \mu(i)x_i - \lambda \sum_{i\neq j} \beta(i, j) x_i x_j  - \Gamma\left( \sum_{i=0}^{N-1}x_i - M\right)^2,
\end{split}
\label{eq:with-penalty-sentences}    
\end{equation}
which can be simplified further by ignoring constant terms to get a quadratic objective:
\begin{equation}
\small
\begin{split}
   \max_{\vec{x} \in \{0,1\}^N} \sum_{i=0}^{N-1}\underbrace{(\mu(i) + 2\Gamma M - \Gamma)}_{\mu_{\Gamma, M}(i)}x_i  - \sum_{i\neq j}  \underbrace{(\lambda\beta(i, j)+\Gamma)}_{{\beta_{\Gamma}(i, j)}} x_i x_j.
\label{eq:last_formulation}   
\end{split}
\end{equation}

\section{Methods}\label{sec:methods}

To generate the optimization problems to be solved, we use articles from the CNN/DailyMail dataset \cite{hermann2015teaching}. This dataset contains just over $300$k unique news articles written by journalists at CNN and the Daily Mail in English. The optimization-problem instances consist of two sets of 10 instances: one set with $N=20$ and a required summary length of $M=8$ and another set with $N=14$ and a required summary length of $M=8$. We use sentence embeddings produced by BERT~\cite{reimers-2019-sentence-bert} to compute similarities, and tf-idf \cite{aizawa2003information} to compute centralities, 
both of which are discussed in detail in Appendix~\ref{sec:appendix-formulation}.%

To quantify the quality of the solution for the optimization problem, we report the approximation ratio given by 
\begin{equation}
    \frac{f_{\text{observed}}-f_{\min}}{f_{\max}-f_{\min}},
\end{equation}
where $f_{\text{observed}}$ is the objective value for the solution produced by a given algorithm, $f_{\max}$ is the maximum value of the objective function and $f_{\min}$ the minimum. The maximum and minimum are computed over all possible in-constraint solutions. For quantum algorithms, $f_{\text{observed}}$ is computed as the average value of the objective function over all in-constraint samples. We additionally report the probability of the solution being in-constraint, which for the experimental results is estimated as the ratio of in-constraint samples to the total number of samples. In practice, the running time of the algorithm scales inversely proportionally to the in-constraint probability due to the need to obtain at least one in-constraint sample.

For unconstrained quantum optimization solvers, the cost function to maximize contains a penalty term with weight $\Gamma$ to constrain the number of sentences in the summary to be $M$ (last term in \eqref{eq:with-penalty-sentences}). The value of this hyperparameter $\Gamma$ was chosen to ensure the value of the cost in \eqref{eq:with-penalty-sentences} corresponding to in-constraint binary strings is greater than the cost corresponding to out-of-constraint binary strings. For each article, after having calculated the similarity and centrality measures, we set $\Gamma = \sum_{i=0}^{N-1} \mu(i) + \lambda \sum_{i\neq j} \beta(i,j)$, where $\lambda=0.075$. See Appendix \ref{appendix:hyperparameter} for additional numerical experiments showing the impact of the value of $\lambda$ on the performance.

We execute the QAOA, XY-QAOA and L-VQE using $14$ and $20$ qubits respectively. We use one layer for QAOA and XY-QAOA ($p=1$) and for L-VQE we use $p=1$ for $14$-qubit problems and $p=2$ for $20$-qubit problems. We optimize the parameters in noiseless simulation and then execute the circuits with optimized parameters on hardware with 2000 shots. We use Qiskit~\cite{aleksandrowicz2019qiskit} for circuit manipulation and noiseless simulation. For the hardware experiments, we transpile and optimize the circuits to H1-1's native gate set using Quantinuum's t$\ket{\text{ket}}$ transpiler \cite{Sivarajah_2020}. For comparison, we also run the quantum algorithms on an emulator provided by Quantinuum, which approximates the noise of H1-1.

For the XY-QAOA circuit, a significant part of the two-qubit gate depth comes from the circuit preparing the initial state, which is a uniform superposition of all in-constraint states (Dicke state). To obtain circuits that are shallow enough to be executable on hardware, we leverage recently developed techniques for the short-depth Dicke-state preparation~\cite{Brtschi2019, 2007.01681, Aktar2021}. Specifically, we use the divide-and-conquer approach~\cite{Aktar2021} to generate circuits targeting a device with an all-to-all connectivity.

\begin{figure}[t]
    \centering
    \includegraphics[width=\textwidth]{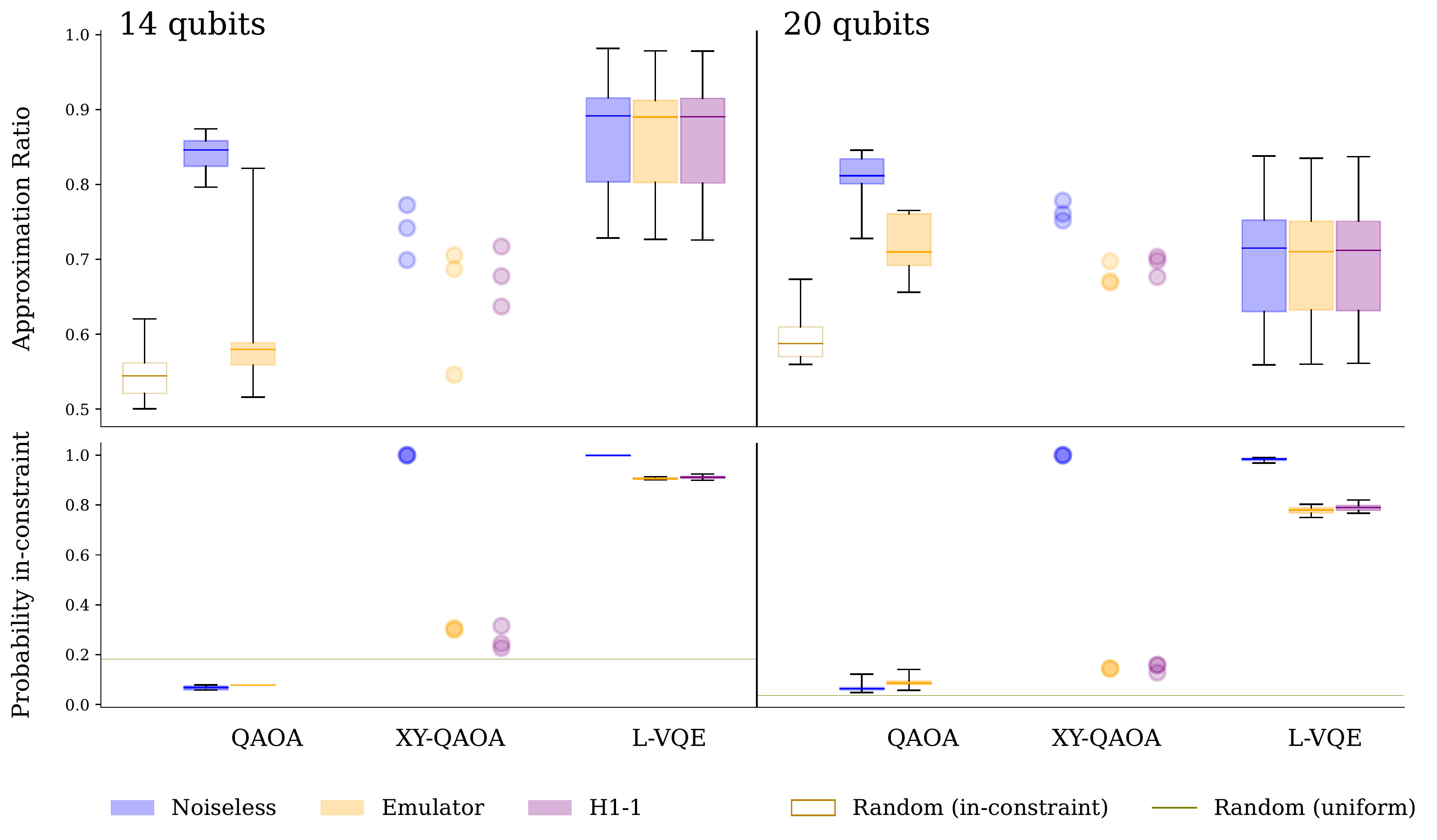}
    \caption{The approximation ratios (top) and in-constraint probabilities (bottom) obtained in the optimization of the instances on $14$ (left) and $20$ (right) qubits using different quantum optimization algorithms executed in a noiseless simulator, the Quantinuum H1-1 machine and its emulator. In each box plot, the box shows quartiles, the line is the median and the whiskers show minimum and maximum. Each plot is showing the statistics over 10 problem instances, with the exception of XY-QAOA where only three 14-qubit and three 20-qubit runs have been executed due to high circuit depth. We observe that all results significantly improve on random guess. The in-constraint probability of QAOA is below random guess for 14 qubits due to the choice of parameters; see discussion in Section~\ref{sec:qaoa_tradeoff}.}
    \label{fig:approximation-ratio}
\end{figure}

\section{Results}
\label{sec:expResults}
We now present experimental results obtained in simulation and on the Quantinuum H1-1 quantum processor. We show the largest to date demonstration of constrained quantum optimization on gate-based quantum computers, using up to 20 qubits and 765 native two-qubit gates (see Section~\ref{sec:related} for a review of previous demonstrations). Note that in the results presented below, we do not perform any error mitigation for the results obtained from hardware or noisy simulations. \rev{The specifications of the H1-1 processor are given in Appendix~\ref{sec:h1-1_details}.}

As previously discussed, we use three quantum optimization algorithms to solve a constrained-optimization problem with the eventual goal of generating document summaries. The optimization algorithms we use are QAOA, L-VQE and XY-QAOA. See Sec.~\ref{sec:problem} for the definitions and discussion of the algorithms, and Appendix~\ref{appendix:numerical_studies} for the implementation details. All the statistics we report are computed over 10 problem instances for each number of qubits, with the exception of XY-QAOA on H1-1 where only three 14-qubit and three 20-qubit instances are solved due to the high circuit depth and correspondingly high running time on trapped-ion hardware. ``Random'' and ``Random in-constraint'' always refers to statistics computed over all binary strings and all in-constraint binary strings respectively. This is equivalent to computing them with respect to uniform random distribution over corresponding sets.

\begin{figure}[ht]
    \centering
    \includegraphics[width=\textwidth]{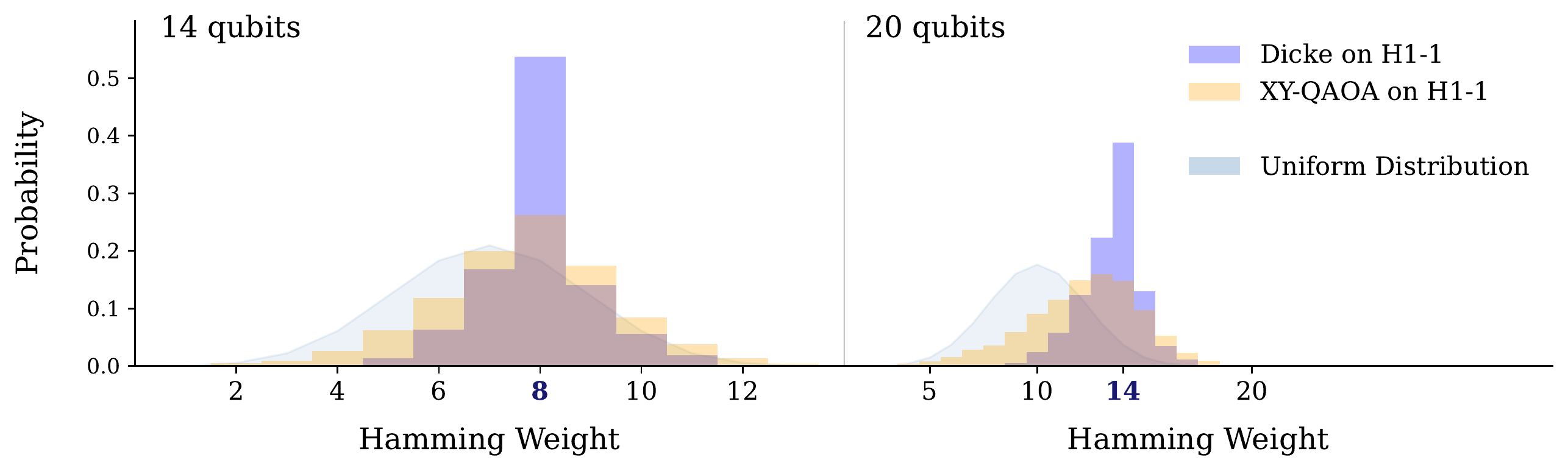}
    \caption{Hamming weight of the bitstrings sampled from the initial uniform superposition of in-constraint states (Dicke state) and a full XY-QAOA circuit. The in-constraint Hamming weight is highlighted in bold. Unlike the QAOA or L-VQE, in XY-QAOA a significant amount of noise is incurred in the initial state preparation step. Observe that the initial Dicke state, due to hardware noise, includes out-of-constraint states.}
    \label{fig:hamming-weight-dicke}
\end{figure}

\begin{figure}[ht]
    \centering
    \includegraphics[width=\textwidth]{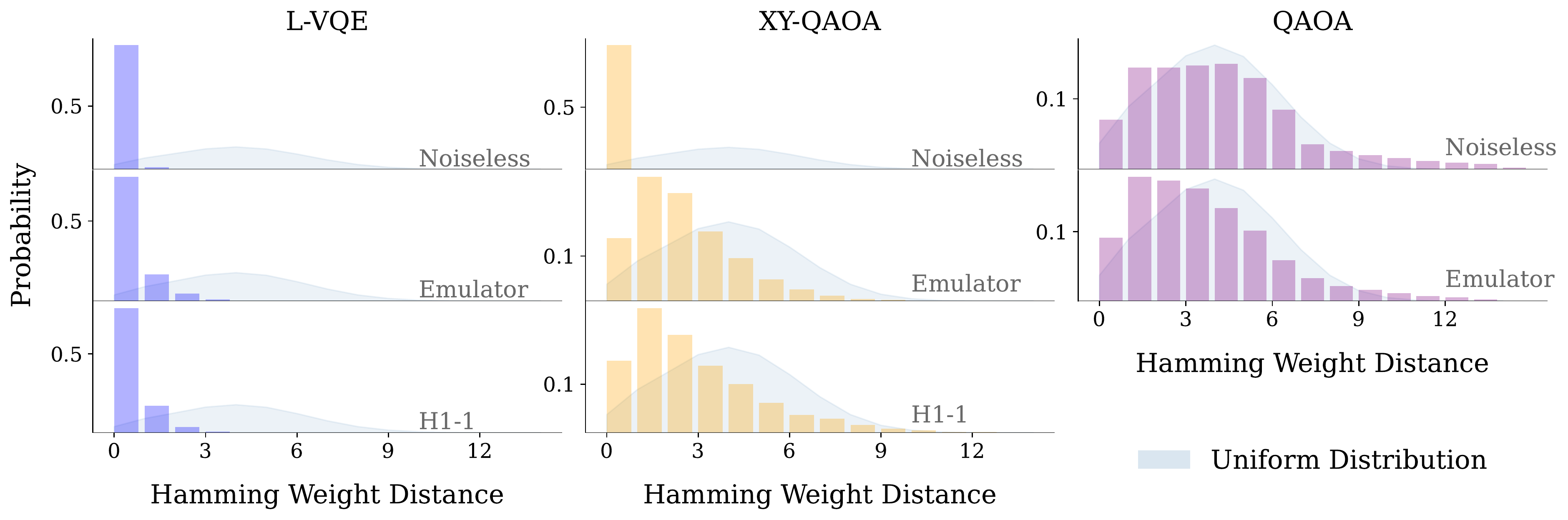}
    \caption{The probability of sampling bitstrings with a given Hamming weight distance to the in-constraint subspace for 20 qubits. The Hamming weight distance is given by
    $|\mbox{wt}(x)-M|$,
    where 
    $\mbox{wt}(x)$ 
    is the Hamming weight of $x$ and $M$ is the constraint value. The shaded background represents the distribution of Hamming weight distance for random bitstrings sampled from a uniform distribution. Note that the overlap with the in-constraint subspace is lower on hardware due to the presence of noise.}
    \label{fig:hamming-distance-20}
\end{figure}

\subsection{Experiments on Hardware}

In Figure \ref{fig:approximation-ratio}, we present the approximation ratios and the in-constraint probabilities for the three algorithms obtained from the execution in a noiseless simulator, with approximated noise in the emulator of H1-1 and on the real H1-1 device. For comparison, we also present the expected approximation ratio of a random feasible solution. We observe that all the approximation ratios obtained, including those from largest circuit executions on hardware, significantly improve upon random guess. Additionally, we observe that the results on hardware are on average at least as good as those obtained from the emulator, making us confident that the emulator gives a good lower bound on the solution quality that can be expected on hardware.

Examining Figure~\ref{fig:approximation-ratio} makes apparent the relative advantages and limitations of the three algorithms we consider. We begin by noting that QAOA gives a relatively high approximation ratio, but a very low in-constraint probability. This is due to the QAOA parameters being chosen to trade-off the two metrics of success; we discuss this issue in detail in Section~\ref{sec:qaoa_tradeoff} below. Here we simply note that the low in-constraint probability of QAOA makes the high approximation ratio less significant. Combined with the complexity of parameter setting arising from the trade-off, this means that QAOA is not a good algorithm for the problem we consider here despite having a relatively high approximation ratio. This trade-off is avoided by L-VQE and XY-QAOA in two different ways. 

L-VQE uses a very expressive parameterized circuit that can in principle solve the problem exactly with no two-qubit gates, just by optimizing the parameters of the initial single-qubit-gate layer $V(\vec{\theta}_0)$. At the same time, the expressiveness of L-VQE circuit makes the parameters hard to optimize, both due to their high number and due to the  gradients vanishing as the number of qubits grows in some cases~\cite{McClean2018,Holmes2022}. As we consider modestly sized problems in this work, we are able to optimize the parameters and obtain solutions with very high approximation ratio and in-constraint probability. However, as the number of qubits grows, this will become increasingly infeasible.

XY-QAOA natively encodes the constraints by restricting the quantum evolution to the subspace equal to the span of the computational basis states of Hamming weight $M$. This leads to an in-constraint probability of one and a high approximation ratio, if no noise is present. At the same time, XY-QAOA requires deeper circuits as compared to QAOA and L-VQE. This is due to the higher gate count cost of the XY-QAOA mixer operator and the initial state preparation.  \rev{The two-qubit gate counts and two-qubit gate depths of the executed circuits are given in Table~\ref{tab:gate_counts}.}

The unconstrained QAOA uses a mixer unitary which is a product of single-qubit rotations, i.e., $U_M(\beta_j) = \prod_{k=1}^N e^{-i\beta_j \xgate_k}$, where $N$ is the number of qubits. For most near-term circuit architectures, including H1-1, the single qubit gates are relatively less noisy than the two-qubit entangling gates~\cite{Pino_2021, ryananderson_2021}. As a result, the mixer unitary does not add much noise to the evolution. Each time-step of QAOA, therefore, has at most $N(N-1)/2$ entangling gates, all of which come from the pairwise interaction in the problem Hamiltonian. On the other hand, XY-QAOA uses a more complex mixer operator, which preserves the Hamming weight of the states it acts on. This operator is defined as: $U_{M}^{XY}(\beta_j) = \prod_{k=1}^N e^{-i\frac{\beta_j}{2}(\xgate_k\xgate_{k+1}+\ygate_k\ygate_{k+1})}$. It adds additional $O(N)$ gates in each layer of QAOA circuit as it requires entangling gates on all adjacent pairs.

The implementation of XY-QAOA requires preparing an initial state, which is a uniform superposition of all in-constraint states (Dicke state). This state is non-trivial to implement, as it requires a circuit with $O(N)$ two-qubit gate depth~\cite{Brtschi2019}. This is in sharp contrast with the initial state used by QAOA or L-VQE, which requires no two-qubit gates to prepare. Therefore in XY-QAOA a significant cost is incurred before any optimization is performed. Figure~\ref{fig:hamming-weight-dicke} shows this by plotting the probabilities of sampling bitstrings with different Hamming weights from the output of the initial state preparation circuit and the full XY-QAOA circuit on the H1-1 device. Unlike the noiseless case, the initial state is not fully contained in the in-constraint subspace. At the same time, the noise is sufficiently low so that the XY-QAOA output distribution is still concentrated on the in-constraint subspace. Improved hardware fidelity would lead to more accurate initial state preparation and higher overall in-constraint probability.

Finally, we examine the impact of noise on the in-constraint probability of the output of all three algorithms.  As discussed above, for our problem of text summarization, it is crucial that the Hamming weight of the output solution is constrained. Both the algorithm design and the hardware noise affect the in-constraint probability. In Figure~\ref{fig:hamming-distance-20}, we visualize this behavior for 20-qubit instances, with the analogous figure for 14 qubits given in Appendix~\ref{appendix:numerical_studies}. Concretely, we plot the probability of obtaining a bitstring $\bm{x}$ with Hamming distance $k$ to in-constraint bitstrings for each $k \in \{0, \dots, N-M\}$. Hamming distance $k$ means that at least $k$ bitflips are required to transform one bitstring into the other. We can see that as the noise is added, the in-constraint probability decreases and the output begins to include out-of-constraint bitstrings. Note that for the short L-VQE circuits, the amount of noise accumulated during the circuit execution is small, and only the bitstrings that are one or two bitflips away are included in the output. On the other hand, for deeper XY-QAOA circuits, bitstrings as far as 10 bitflips away are included. For QAOA, even the noiseless output includes primarily out-of-constraint bitstrings due to the choice of parameters.

\begin{figure}[t]
    \centering
    \includegraphics[width=\textwidth]{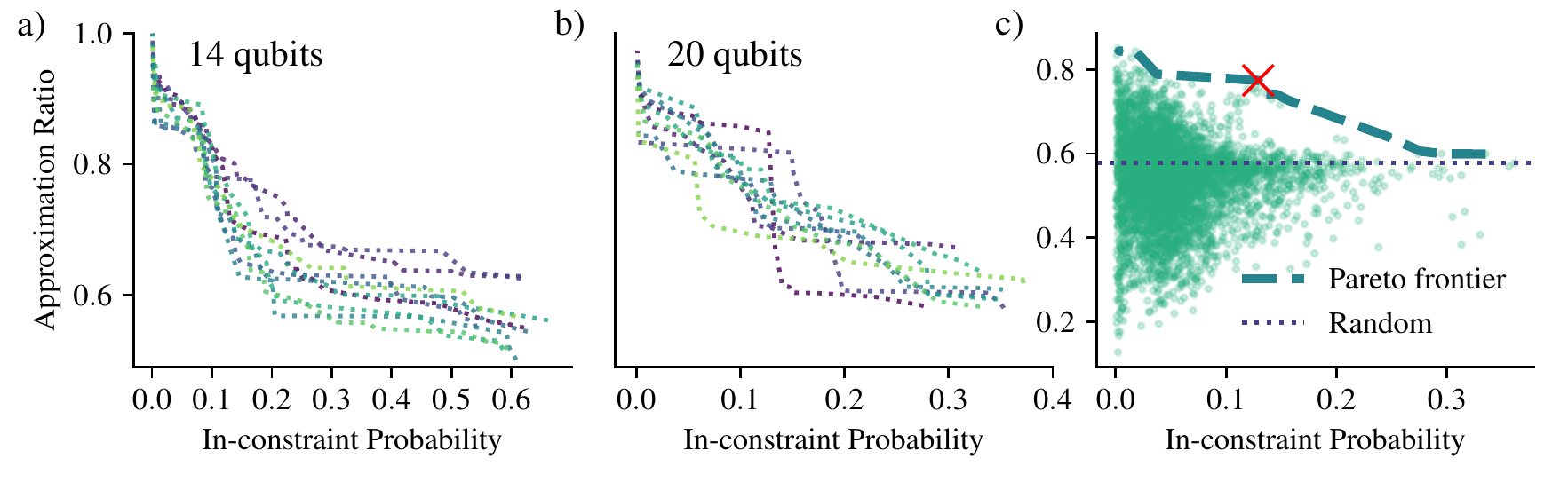}
    \caption{Pareto frontier of QAOA parameters with respect to approximation ratio and the in-constraint probability for 14 (a) and 20 (b) qubits. Each color plots a different problem instance for a given number of qubits.  When solving a constrained-optimization problem using QAOA, the parameter optimization strategy has to be tuned to arrive at the desired point on the Pareto frontier. Directly optimizing the objective \eqref{eq:with-penalty-sentences} would lead to choosing the righmost point of the frontier, which in the case of the 20 qubit instance presented in (c) corresponds to approximation ratio equal to that of random guess. In (c), each dot corresponds to one value of $\vec{\beta}, \vec{\gamma}$ from a grid search in parameter space. Thick dashed line shows the Pareto frontier and the thin dotted line marks the approximation ratio of a random solution. The red {\color{red} $\times$} marker indicates the parameters chosen for the experiments shown above.}
    \label{fig:trade_off}
\end{figure}

\begin{table}[b]
    \centering
    \begin{tabular}{c|c|c|c|c|c}
         \multirow{2}{*}{\rev{Algorithm}} & \multirow{2}{*}{\rev{\# qubits}} & \multicolumn{2}{c|}{\rev{Initial State}} & \multicolumn{2}{c}{\rev{Full Circuit}}  \\
         & & \rev{2Q depth} & \rev{2Q count} &  \rev{2Q depth} & \rev{2Q count}\\
         \hline
         \multirow{2}{*}{\rev{L-VQE}} & \rev{14} & \rev{0} & \rev{0} & \rev{4} & \rev{26} \\
         & \rev{20} & \rev{0} & \rev{0} & \rev{8} & \rev{76} \\
         \hline
         \multirow{2}{*}{\rev{QAOA}} & \rev{14} & \rev{0} & \rev{0} & \rev{50} & \rev{182} \\
         & \rev{20} & \rev{0} & \rev{0} & \rev{74} & \rev{380} \\
         \hline
         \multirow{2}{*}{\rev{XY-QAOA}} & \rev{14} & \rev{68} & \rev{185} & \rev{114} & \rev{393} \\
         & \rev{20} & \rev{95} & \rev{347} & \rev{159} & \rev{765} \\
    \end{tabular}
    \caption{\rev{Two-qubit gate depths (2Q depth) and two-qubit gate counts (2Q count) of the circuits executed on emulator and on the H1-1 device. The circuits were optimized using pytket~\cite{Sivarajah2020}, and the optimized numbers are reported. Initial state preparation for QAOA and L-VQE requires no two-qubit gates.}}
    \label{tab:gate_counts}
\end{table}

\subsection{Quantum Circuit Needs to Preserve Constraints}\label{sec:qaoa_tradeoff}

As discussed above, the output distribution of QAOA circuits has a relatively small overlap with the in-constraint space even in the absence of noise. This is due to the trade-off between the in-constraint probability and the approximation ratio, which is implicit in the choice of the parameter optimization strategy. This trade-off is an important weakness of unconstrained quantum algorithms applied to constrained problems, necessitating the development and implementation of quantum algorithms that natively preserve constraints. In the QAOA experiments presented in Figure~\ref{fig:approximation-ratio}, we choose the parameters with the goal of increasing the approximation ratio of in-constraint solutions at the expense of reduced in-constraint probability. We now examine this trade-off numerically.

We perform a grid search over the parameter space $\vec{\beta}, \vec{\gamma}$ of a single-layer QAOA. In Figure~\ref{fig:trade_off} we plot the Pareto frontier with respect to approximation ratio and in-constraint probability. Specifically, we plot the approximation ratios and in-constraint probabilities for QAOA with parameter values $\vec{\beta}, \vec{\gamma}$ such that there do not exist parameters $\hat{\vec{\beta}}, \hat{\vec{\gamma}}$, which improve either of the metrics without decreasing the other. We plot such frontiers for all 20 instances considered.

This trade-off behavior is not specific to QAOA and indeed applies to all unconstrained quantum algorithms that are not expressive enough to solve the problem exactly. Examining the Pareto frontiers makes clear the challenge of choosing parameters for such algorithms. Whereas in this work we perform the full grid search and we may actually choose any point on the frontier for execution on hardware, in practice an objective function must be carefully designed to optimally trade-off the two objectives. This is hard in general. As an example, optimizing parameters with respect to \eqref{eq:with-penalty-sentences} would lead to prioritizing the in-constraint probability. Figure~\ref{fig:trade_off}(c) shows an example of an instance where this leads to approximation ratio equal to that of random guess.

One potential solution to this challenge is using a circuit that is sufficiently expressive to solve the problem exactly, such as the circuit used in L-VQE. However, such circuits by necessity would have many parameters that are hard to optimize. Additionally, in many cases they would suffer from gradients of the objective vanishing exponentially with the number of qubits (``barren plateaus'')~\cite{Holmes2022}, making optimization impossible. Therefore the most practical solution is encoding constraints directly into the quantum circuit, as is the case for XY-QAOA.

\bgroup
\def\arraystretch{1.6}%
\setlength\tabcolsep{4pt}

\begin{table}[]
    \centering
    \begin{tabular}{c|c|c|c|c|c|c|c|c|c}
        & Ref & Year & Hardware  & Connectivity & $N$ & $p$ & 2q-gate depth & Error Mit. & H.W.C.  \\ \hline
        \multirow{4}{*}{\rotatebox[origin=c]{90}{Unconstrained}}
        & \cite{otterbach2017} & 2017 & superconducting & nearest neighbor &19  & 1 & 6 & No & - \\ 
        & \cite{Lacroix_2020} & 2020 &  superconducting & nearest neighbor & 7 & 6 & 42 & No & - \\ 
        & \cite{harrigan2021quantum} & 2021 & superconducting & nearest neighbor & 23 & 5 & 40\footnote{\label{footnote:estimated}Depth estimated from the circuit description in the paper.} & Yes & - \\ 
        & \cite{harrigan2021quantum} & 2021 & superconducting & nearest neighbor & 17 & 3 & 153  & Yes & - \\
        \hline
        \multirow{4}{*}{\rotatebox[origin=c]{90}{Hard constraint}}
        & \cite{baker2022} & 2022 & superconducting & nearest neighbor & 3 & 5 & - & No & 2 \\ 
        & \cite{baker2022} & 2022 & trapped ion & all-to-all & 3 & 4 & - & No & 2 \\ 
        & This work & 2022 & trapped ion & all-to-all & 14 & 1 & 114  & No & 8\\ 
        & This work & 2022 & trapped ion & all-to-all & 20 & 1 & $\boldsymbol{159}$ & No & 14

    \end{tabular}
    \caption{Comparison of the hardware demonstrations shown in this work with previous hardware demonstrations of QAOA on gate-based devices. We only report experiments that  outperformed random guess and either are similar in scale to ours in terms of circuit size or utilized hard constraints. All hard constraints are on the Hamming weight (H.W.C. in the table). The two-qubit-gate depth is missing from references where the depth was not reported and could not be estimated. We also report the topology of two-qubit interactions available o the hardware (``Connectivity'' in the table) and whether error-mitigation techniques were applied (``Error Mit.'').}
    \label{tab:previous_results}
\end{table}

\section{Related Work \label{sec:related}}

In this work, we show the largest demonstration to date of constrained optimization on a gate-based quantum computer. We now briefly review the state-of-the-art, which we summarize in Table \ref{tab:previous_results}. We include the unconstrained-optimization demonstrations for completeness, and emphasize that the circuits used in this work are deeper than any quantum optimization circuits executed previously for any problem.

There have been various quantum hardware demonstrations of QAOA applied to unconstrained-optimization problems. Using Google's ``Sycamore'' superconducting processor, Harrigan \etal\cite{harrigan2021quantum} ran QAOA to find the ground states of Ising models that mapped to the 23-qubit Sycamore  topology and Sherrington--Kirkpatrick models with 17 vertices. Additionally, they solved MaxCut on 3-regular graphs with up to 22 vertices. Otterbach \etal solved MaxCut on a 19-vertex graph that obeys the hardware topology of Rigetti's 19-qubit ``Acorn'' superconducting processor \cite{otterbach2017}. Lacroix \etal \cite{Lacroix_2020} executed QAOA, up to $p=6$, on a superconducting gate-based quantum computer to solve the exact-cover problem on at most seven vertices. The device supported two-qubit controlled arbitrary-phase gates. This enhanced gate set resulted in a two-qubit-gate depth of 42. 
There have been other hardware demonstrations of solving unconstrained-optimization problems, with either lower qubit counts or shallower circuits~\cite{Qiang_2018, Willsch_2020, Abrams_2020, Bengtsson_2020, Earnest_2021, Santra_2022, shaydulin2021error, 2204.05852, Shaydulin2019, UshijimaMwesigwa2021}.

An important step in implementing XY-QAOA is the preparation of Dicke states. Aktar \etal\cite{Aktar2021}, who developed the divide-and-conquer Dicke-state preparation approach used in our experiments, implemented their algorithm using up to six qubits on two IBM Q devices. 

To the authors' knowledge, there has only be one other demonstration of QAOA with constrained mixers and Dicke-state initialization on quantum hardware. This was done by Baker and Radha \cite{baker2022} who solved problems using at most five qubits and $p=5$. They executed circuits using both the XY complete-graph mixer with Dicke-state initialization and the XY ring mixer with initialization to a random in-constraint state. They utilized the Rigetti ``Aspen-10'' superconducting processor, five IBM Q superconducting processors, and the 11-qubit IonQ trapped-ion device. They reported that QAOA beat, with both the XY complete-graph mixer and XY ring mixer, random guess for up to three qubits and $p=5$ on the Rigetti device. Lastly, their results on IonQ, for the XY ring mixer, beat random guess for up to three qubits and $p=4$.

While there have been other hardware demonstrations of QAOA using alternative mixers to the transverse field, they were either not applied to hard-constrained problems or did not use an in-constraint initial state. For example, Golden \etal\cite{golden2021} solved six Ising problems on 10 IBM Q backends using QAOA with Grover mixers \cite{Bartschi_2020}. While Grover mixers can be used to incorporate hard constraints, the problems Golden \etal solved were unconstrained. Pelofske \etal\cite{Pelofske_2021} solved five Ising problems, still unconstrained, with QAOA using Grover mixers on seven IBM Q backends, Rigetti's ``Aspen-9'' device, and the 11-qubit IonQ device. The instances required fewer than seven qubits. For a protein folding problem, Fingerhuth \etal\cite{Fingerhuth2018} executed QAOA with an XY mixer using four qubits and $p=1$ on Rigetti's Acorn device.  However, the initial state was a uniform superposition over all bitstrings, and thus the Hamming weight constraint was not obeyed. 

Lastly, there have been very large-scale demonstrations of QAOA on analog quantum simulators. First, Pagano \etal \cite{Pagano_2020} demonstrated QAOA on a trapped-ion analog quantum simulator using up to 40 qubits at $p=1$ and 20 qubits at $p=2$ on unconstrained problems. Second Ebadi \etal\cite{ebadi2022quantum} show an application of an analog quantum simulator, using QAOA, to maximum independent set (MIS) problems on graphs with up to 179 vertices. This variant of QAOA was performed by controlling the timing of global laser pulses applied to a 2D array of 289 cold atoms. The global pulses induce Rydberg excitations, which result in a blockade effect that ensures that only independent sets are sampled. However, since these analog devices do not implement universal gate sets, the results are not directly comparable to ours.

\section{Discussion}
\label{sec:Discussion}
In this work, we present the largest to date demonstration of constrained optimization on quantum hardware. Our results demonstrate how algorithmic and hardware progress are bringing the prospect of quantum advantage in constrained optimization closer, which can be leveraged in many industries including finance \cite{https://doi.org/10.48550/arxiv.2201.02773,9643469}. 

In our experiments on the $20$-qubit Quantinuum H1-1 system, we observe that XY-QAOA with up to $20$ qubits provides results that are significantly better than random guess despite the circuit depth exceeding $100$ two-qubit gates. This progress can be clearly observed by comparing the size and the complexity of the circuits used in our experiments to the previous results discussed in Section~\ref{sec:related} above. The results we present here were obtained with no error mitigation, which is not the case for many of the previous demonstrations. Our execution of complex circuits for constrained optimization benefits from the underlying hardware's  all-to-all connectivity, as the circuit depth would increase significantly if the circuit had to be compiled to a nearest-neighbor architecture.

We additionally show the necessity of embedding the constraints directly into the quantum circuit being used. If the circuit does not preserve the constraints, the in-constraint probability and the quality of the in-constraint solution have to be traded-off against each other. This trade-off is hard to do in general. This observation further motivates our investigation of XY-QAOA on H1-1, and gives additional weight to our results. At the same time, we show that further advances are needed to reduce the hardware requirements of implementing such circuits and improve the fidelities of the hardware..

\section*{Acknowledgements}

The authors thank Tony Uttley, Brian Neyenhuis, Jenni Strabley and the whole Quantinuum team for their support and feedback, and especially for providing us preview access to the Quantinuum H1-1 upgraded to 20 qubits. The authors thank Andreas B{\"{a}}rtschi and Stephan Eidenbenz for the helpful discussions on the Dicke-state preparation. Additionally, the authors appreciate the support of their FLARE colleagues at JPMorgan Chase. P.N. acknowledges funding by the DoE ASCR Accelerated Research in Quantum Computing program (award No.~DE-SC0020312), DoE QSA, NSL QLCI (award No.~OMA-2120757), NSF PFCQC program, DoE ASCR Quantum Testbed Pathfinder program (award No.~DE-SC0019040), U.S.~Department of Energy Award No.~DE-SC0019499, AFOSR, ARO MURI, AFOSR MURI, and DARPA SAVaNT ADVENT.

\section*{Author Contributions Statement}

P.N., R.Y. devised and implemented the optimization formulation. R.S., R.Y. implemented the quantum algorithms.  P.N., R.S., R.Y., D.H. performed the experiments. P.N., R.S., R.Y. analyzed the data. M.P. supervised the project.  All authors contributed to the technical discussions, evaluation of the results, and the writing of the manuscript.

\section*{Additional Information}
The authors declare no competing interests. 

\section*{Data Availability Statement}
The data presented in the paper is available at \url{https://doi.org/10.5281/zenodo.6819861}

\section*{Disclaimer}
This paper was prepared for information purposes by the Future Lab for Applied Research and Engineering (FLARE) group of JPMorgan Chase Bank, N.A.. This paper is not a product of the Research Department of JPMorgan Chase Bank, N.A. or its affiliates. Neither JPMorgan Chase Bank, N.A. nor any of its affiliates make any explicit or implied representation or warranty and none of them accept any liability in connection with this paper, including, but limited to, the completeness, accuracy, reliability of information contained herein and the potential legal, compliance, tax or accounting effects thereof. This document is not intended as investment research or investment advice, or a recommendation, offer or solicitation for the purchase or sale of any security, financial instrument, financial product or service, or to be used in any way for evaluating the merits of participating in any transaction.

\bibliography{citations}

\clearpage

\pagenumbering{arabic}%
\renewcommand*{\thepage}{\arabic{page}}
\appendix 

\section*{Appendix}

\medskip

\section{Details of the Optimization Problem Formulation}\label{sec:appendix-formulation}
The formulation of extractive summarization as an optimization problem requires a meaningful measure of centrality of sentences and similarity between two sentences. To calculate similarity, we use vector embeddings obtained from a neural language model  \cite{reimers-2019-sentence-bert}, specifically the Bidirectional Encoder Representations from Transformers (BERT) \cite{devlin2018bert}. Neural Transformers \cite{vaswani2017attention} have resulted in many recent successes in various natural language processing tasks \cite{wang2019language}. We measure sentence similarity by computing the \emph{cosine similarity} of BERT embeddings defined as
\begin{equation} \frac{\vec{v}_i\cdot\vec{v}_j}{\lVert\vec{v}_i\rVert_2\lVert\vec{v}_j\rVert_2},
\end{equation}
where $\vec{v}_j \cdot \vec{v}_i$ denotes the usual dot product, and $\lVert\cdot\rVert_2$ the $\ell_2$ norm. We refer the reader to the work done by Achananuparp \etal \cite{achananuparp2008evaluation} for various techniques used to measure similarity.

For a measure of centrality, we use the tf-idf statistic \cite{aizawa2003information, zheng2019sentence}. In contrast to embeddings obtained from pre-trained language models like BERT, \textit{term frequency–inverse document frequency} (tf-idf) is a technique of measuring the ``salience'' or ``importance'' of a word in a document by counting its occurence in the entire corpus. The tf-idf measure of a word is computed by multiplying a measure of term frequency (tf) with a measure of inverse document frequency (idf). The tf is the proportion of a document, $D$, that contains the word $w$. The idf scales down the raw proportion based on the how frequently the word occurs across all documents. Mathematically, for a word $w$ appearing in a sentence $s$ of a document $D$,

\begin{equation}
\small 
    \text{tf-idf}(w, S, D) := \underbrace{\left(\frac{f_{w, S}}{\sum_{w'\in S} f_{w',S}}\right)}_{\rm tf(w,S)} \times  \underbrace{\left(\log \frac{N}{|\{S \in D: w \in S\}|}\right)}_{\rm idf(w,D)},
    \label{eq:tf-idf-words}
\end{equation}
where $f_{w, S}$ is the frequency of word $w$ in $S$ and $N$ is the number of unique words in the document.

While the measure in \eqref{eq:tf-idf-words} was defined for words, we need a similar measure defined for sentences. To define a tf-idf-based measure for sentences, we take the mean of the tf-idf values of the words in the sentence:
\begin{equation}
    \overline{\text{tf-idf}}(S) := \frac{1}{n_S}\sum_{w \in S} \text{tf-idf}(w, S, D)
\label{eq:tfidf}    
\end{equation}
where $n_S$ is the length of the sentence $S$.

\section{Details of Numerical Studies}
\label{appendix:numerical_studies}

For each instance of the optimization problem, we evaluate QAOA for each of the values in the grid using $1000$ shots. In order to decide the best parameters for each optimization instance, we impose a threshold on the QAOA in-constraint ratio to be higher than $0.06$ and we selected the $\gamma, \beta$ that maximize the expected approximation ratio. For all the instances, the value obtained was higher than the expected approximation ratio of a random feasible solution.

For L-VQE and XY-QAOA the parameters are optimized by running COBYLA~\cite{powell1994direct,powell1998direct} from a fixed number of randomly chosen initial points. The number of initial points is 20 for L-VQE with 14 qubits, 5 for L-VQE with 5 qubits and 10 for XY-QAOA for all qubit counts. We choose the parameters giving the best approximation ratio.
n.

\begin{figure*}[!ht]
    \centering
    \includegraphics[width=\textwidth]{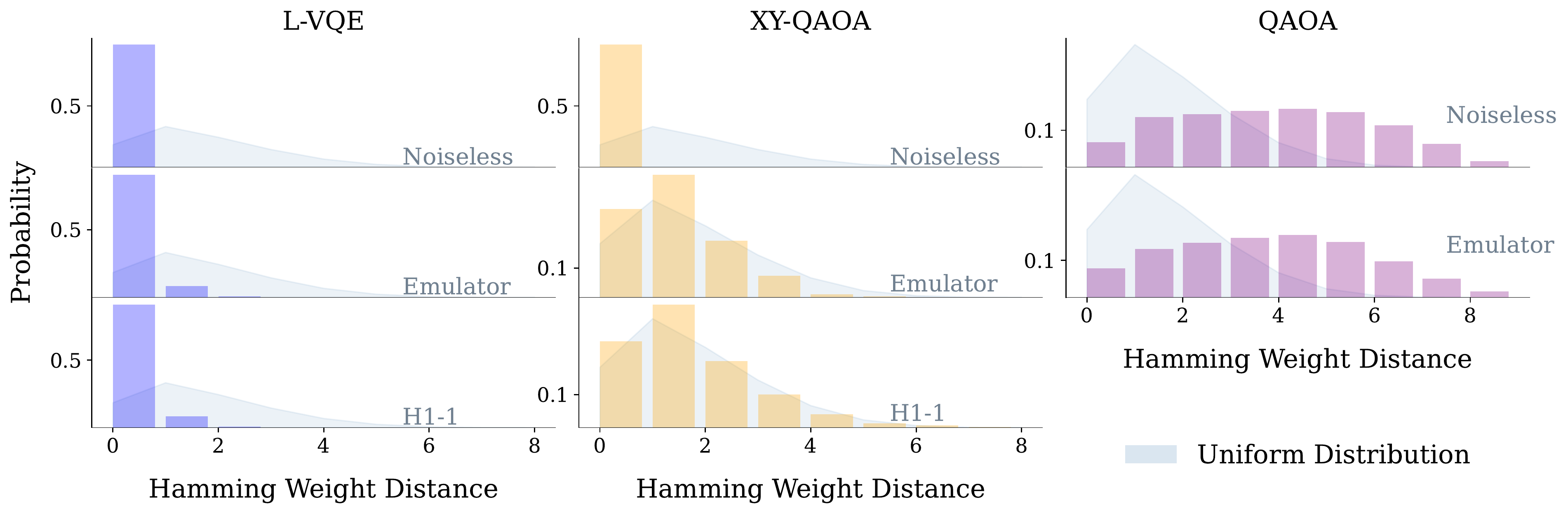}
    \caption{The probability of sampling bitstrings with a given Hamming weight distance to the in-constraint subspace for 14 qubits. The Hamming weight distance is given by 
    $|\mbox{wt}(x) - M|$, 
    where 
    $\mbox{wt}(x)$
    is the Hamming weight of $x$ and $M$ is the constraint value. The shaded background represents the distribution of Hamming weight distance for random bitstrings sampled from a uniform distribution. Note that the overlap with the in-constraint subspace is lower on hardware due to the presence of noise. For QAOA, the in-constraint probability is lower than that of random guess due to parameter choice as we fix the target in-constraint probability to be 0.06.}
    \label{fig:hamming-distance-14}
\end{figure*}

In the main text, we present the distribution of Hamming weight distances for optimization instances on 20 qubits. For completeness, we present the same distribution for 14 qubits in Figure \ref{fig:hamming-distance-14}. As it happens with $20$ qubits, for algorithms with shallow circuits (L-VQE) the ideal distribution is concentrated around zero even when executed on hardware. Whereas 
XY-QAOA, an algorithm that natively preserve cardinality, when executed on noiseless simulation produces a distribution concentrated at $k=0$, when it is executed on the emulator and real hardware, it produces a much wider distribution and it is more similar to the random distribution. Notably, for QAOA, both in noiseless simulator and in emulator, the distribution is concentrated far away from $k=0$ and it is even concentrated at a $k$ higher that the center of the random distribution. %

\section{\rev{Details of the H1-1 Quantum Processor}}\label{sec:h1-1_details}

\rev{The Quantinuum H1-1 Quantum Processor uses quantum charge-couple device architecture with five parallel gate zones in a linear trap. The quantum states are stored in hyperfine states of twenty \textsuperscript{171}Yb\textsuperscript{+} atoms. All-to-all connectivity is implemented by rearranging of the physical location of qubits, which introduces a negligible amount of error. Typical single-qubit gate infidelity is $5\times 10^{-5}$ and typical two-qubit gate infidelity is $3\times 10^{-3}$. Typical error rate of state preparation and measurement is $3\times 10^{-3}$. Memory error per qubit at average depth-1 circuit (``idle error'') is $4\times 10^{-4}$. Additional details are available in Ref.~\cite{h11datasheet}}

\section{Evaluation of Optimization-Generated Summaries}

A popular evaluation metric used for summarization tasks is the Recall-Oriented Understudy for Gisting Evaluation (ROUGE) metric \cite{lin2004rouge}. We calculate F1 score of the three variants of ROUGE, namely ROUGE-1-F, ROUGE-2-F and ROUGE-L-F, which measure unigram-similarity, bigram similarity and longest-common subsequence, respectively. 

Each of the optimization instances discussed in the main text corresponds to an article from the CNN/DailyMail dataset, consisting of 14 or 20 sentences, with the linear and quadratic coefficients in the optimization objective generated using centrality and similarity measures discussed in Sec \ref{sec:appendix-formulation}. The predicted summaries consist of 8 and 14 sentences respectively. For each of the optimization instances, we calculate the ROUGE metrics with the output distribution of the quantum algorithms, weighted by the probability of measuring an in-constraint bitstring, which corresponds to a summary with the specified length.

In Figure \ref{fig:rouge-experiments}, we present the quality of summaries generated by the quantum algorithms discussed in the main text, in terms of the three ROUGE metrics. For perspective, we also present the metrics when evaluated against an uniform distribution, and also the metrics for the optimal extractive summary for each article. Note that the ROUGE metrics are obtained by comparing predicted summaries against a human-written ``highlight'' that comes associated with the CNN/DailyMail dataset. The highlights are brief paraphrasings of the article and do not necessarily contain any sentences or verbatim phrases from the article. As a result, even optimal extractive summaries have the median score that is far from the maximal ROUGE score of 1. Furthermore, as can be seen in Figure \ref{fig:rouge-experiments}, the optimal summaries are only slightly better than the summaries generated by sampling from a uniform distribution and the solutions obtained from the optimization algorithms discussed in the main text only perform as well as random summaries. The fact that even noiseless simulations of quantum algorithms fail to consistently outperform random summaries suggests that the usefulness of the optimization framework as a route to summary generation remains inconclusive, at least for the dataset, evaluation criteria and the qubit counts we consider in our work.

\begin{figure}[!ht]
    \centering
    \includegraphics[width=\textwidth]{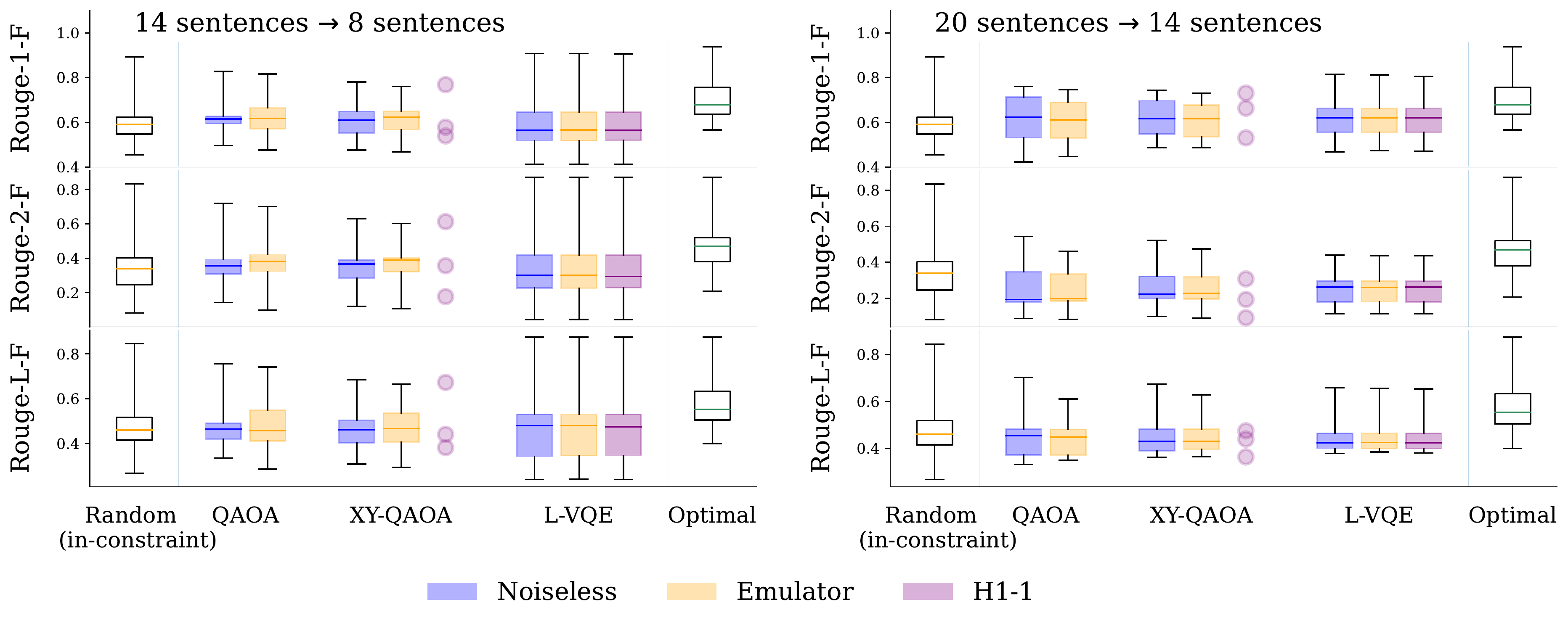}
    \caption{Rouge metrics (1-F, 2-F, and L-F) for extractive summaries generated using the three different optimization algorithms executed in a noiseless simulator, emulator and the real quantum hardware. As helpful guides, we also present the expected ROUGE metric for summaries sampled from the uniform distribution, and also the ROUGE metric for the optimal summary set.}
    \label{fig:rouge-experiments}
\end{figure}

\section{Tuning Hyperparameter $\lambda$}
\label{appendix:hyperparameter}

The complexity of the optimization of  \eqref{eq:summary-optimization} comes from the quadratic terms, which penalize redundancies that are present in the summary. The hyperparameter $\lambda$ controls the strength of the penalty for including redundant sentences. If the value of $\lambda$ is very small, the summary will be informative but potentially redundant. Similarly, if $\lambda$ is too large, the algorithm can pick distinct sentences but with very low information content. 

We study how the penalization for redundancy presented in the summary affects its quality. Specifically, we consider different values of $\lambda$ and compute the average ROUGE metric, for the $10$ optimization instances of 14-sentence articles discussed in the main text. For each article, we generate a set of objective functions \ref{eq:with-penalty-sentences} for $\lambda \in [0, 0.25]$ and find optimal solutions to these problems with brute-force. We plot the ROUGE metrics for the resulting summaries from the optimal solutions in Fig \ref{fig:hyperparameter}. We observe a slight increase in average ROUGE for $\lambda > 0.05$. In the experiments discussed in the main text, we use $\lambda = 0.075$.

\begin{figure}
    \centering
    \includegraphics[width=0.5\linewidth]{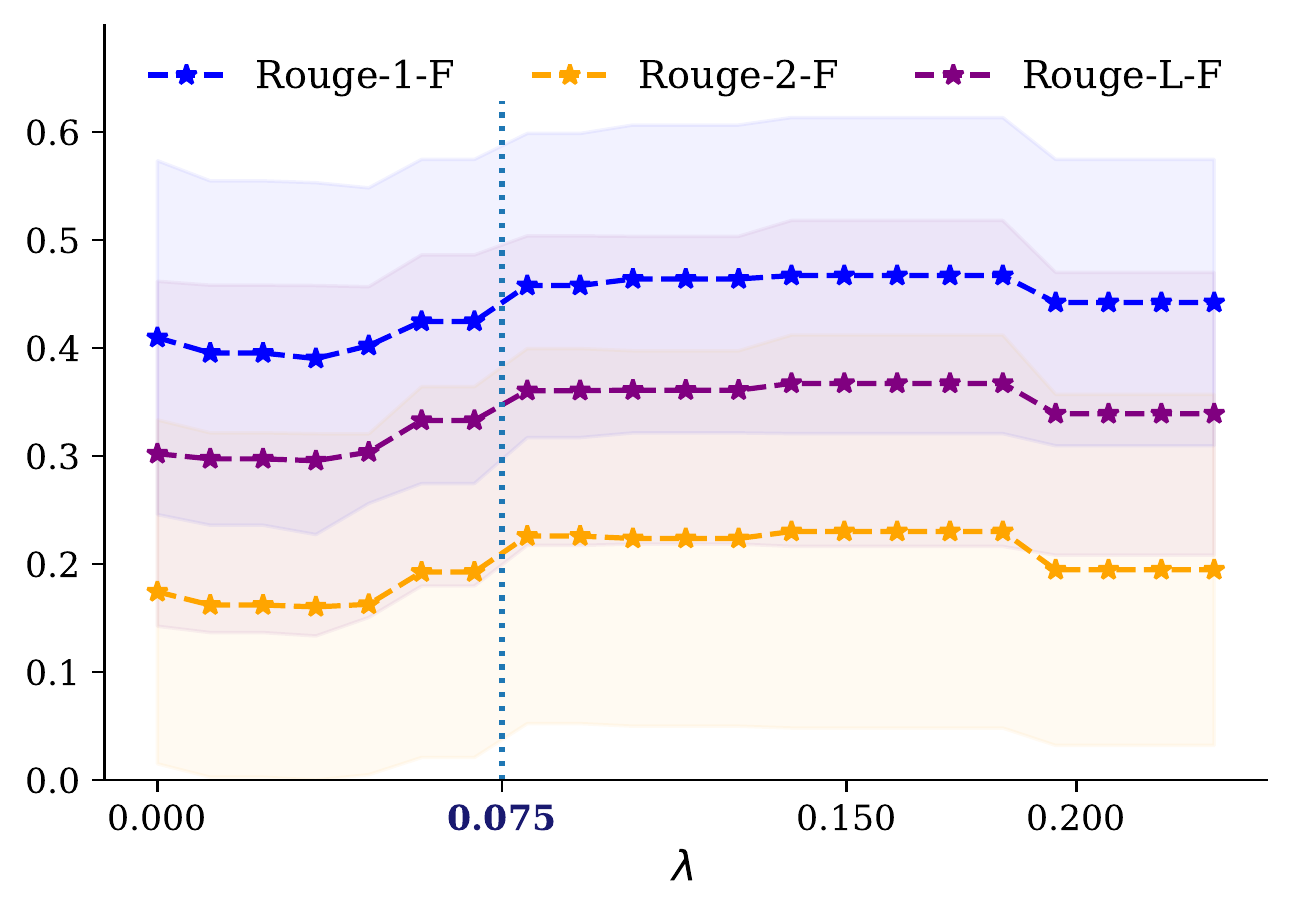}
    \caption{ROUGE metrics for summaries obtained with optimal solutions (obtained via brute-force) for the instances of 14-sentence articles used in the main text, as a function of $\lambda$. The dashed vertical line represents the $\lambda$ used in experiments discussed in the main text. The shaded area represents the standard deviation around the mean.}
    \label{fig:hyperparameter}
\end{figure}
\end{document}